\documentclass[twocolumn,showpacs,preprintnumbers,amsmath,amssymb]{revtex4}

\usepackage{graphicx}
\usepackage{dcolumn}
\usepackage{bm}

\begin{document}

\title{Influence of inelastic relaxation time on intrinsic spin Hall effects
in a disordered two-dimensional electron gas }

\author{Pei Wang and You-Quan Li}
\affiliation{Zhejiang Institute of Modern Physics and Department of Physics,\\
Zhejiang University, Hangzhou 310027, P. R. China}

\date{\today}

\begin{abstract}

The influence of inelastic relaxation time on the intrinsic
spin Hall effects in a disordered
two-dimensional electron gas with Rashba interaction is studied,
which clarifies the controversy of impurity effects in the system.
We reveal that, due to the existence of inelastic scattering,
the spin Hall conductivity does not vanish when the impurity
concentration diminishes to zero
no matter it is non-magnetically or magnetically disordered.
The spin accumulation is evaluated by using the obtained spin Hall conductivity,
and an alternate route is suggested to verify the intrinsic spin Hall
effect by measuring the spin accumulation at different ratios.

\end{abstract}

 \maketitle

\section{Introduction}

Much attention has been paid to the study of  intrinsic spin Hall effects
(ISHE) which is expected to bring in practical applications in spintronics.
In the intrinsic spin Hall effect, the spin current
arises from the spin-orbit-dependent band structure.
Theoretically, ISHE may exist in the p-type semiconductor~\cite{zhang} and
two-dimensional electron gas (2DEG)~\cite{sinova04}.
After Sinova et al.~\cite{sinova04} predicted a universal spin Hall conductivity,
$\sigma^{}_{s H}=e/8\pi$ in clean 2DEG, several
groups~\cite{Mishchenko,Inoue04,Raimondi,Rashba04,Dimitrova}
indicated that an arbitrarily
small impurity concentration would suppress the spin Hall conductivity to zero
due to the vertex corrections.
Rashba~\cite{Rashba04} and Dimitrova~\cite{Dimitrova} proved that the
spin-Hall current was always zero in the non-magnetically disordered system
while
Grimaldi et al.~\cite{Grimaldi} and Krotkov~\cite{Krotkov} noticed that the
spin Hall conductivity is not zero unless it is in the special situation
for a large Fermi circle, quadratic band structure and momentum-independent
Rashba coefficient.
Very recently, Inoue et al.~\cite{Inoue06} and Wang et al.~\cite{Pei} recognized that
the spin Hall conductivity is nonzero in the presence of magnetic impurities.
However, no one can explain why there exists a discontinuous jump in
$\sigma^{}_{sH}$ between the clean limit and a clean system.
Thus there remains a puzzle that why the clean limit of the spin Hall conductivity
does not equal to the one in a clean system.
Opposite to analytical results,
a numerical calculation in a finite 2DEG by Nomura et. al.~\cite{Nomura051}
manifested a robust spin Hall conductivity that falls to zero only
when the inverse of the elastic relaxation time
(elastic lifetime in brevity) is larger than the spin-orbit splitting of the
bands, i.e.,  $1/\tau > \Delta$.
This raises a question that why there is such a controversy
between the numerical calculation and the analytical consequences.
A consistent comprehension of the aforementioned issues becomes an
obligatory issue.

It is known that all the states in two dimensional infinite system
in the presence of disorder are localized~\cite{Anderson791} at
zero temperature. At finite temperature, there exist delocalized
states because of the presence of dephasing. The interference
occurs only inside the decoherence length so that the electronic
conductivity depends on the ratio of elastic to inelastic
lifetimes. Although the importance of dephasing in the electrical
charge transport in 2D systems has been addressed in the weak
localization theory~\cite{Anderson791, Anderson792,Gorkov,
Altshuler, Thouless, Lee, Bergmann}, there has been no discussions
on the role of dephasing in the spin Hall effect. We will show in
the present paper that the dephasing plays a crucial role in the
spin Hall effect, leading to a nonzero conductivity for an
arbitrary impurity concentration.

This paper is organized as follows.
In next section, we give a description of the system  and
make a general formulation by introducing inelastic relaxation time.
In section \ref{sec:nonmagetic}, we investigate the influence on spin
Hall conductivity by nonmagnetic impurities.
In section \ref{sec:magnetic}, we consider magnetic impurities and study
their effects on the spin Hall effect.
In section \ref{sec:impurity}, we plot the curves of the conductivity
versus the ratio of elastic to
inelastic relaxation times and observe their asymptotic behaviors.
We also discuss the corresponding finite temperature behaviors.
In section \ref{sec:accumulation}, we evaluate the spin accumulation
in terms of the spin Hall conductivity we obtained.
Our conclusive remarks are briefly summarized in the last section.

\section{General consideration}

The Hamiltonian for a 2DEG with Rashba spin-orbit coupling is given
by $H=p^2/2m^*+\alpha(\sigma^x p_y-\sigma^y p_x) + V_{dis}$ with
$V_{dis}$ denoting potentials produced by either nonmagnetic
impurities or magnetic impurities. The spin current is defined as
\begin{eqnarray}\nonumber
J_y^z(\textbf{p})=&&\frac{1}{4}(v_y \sigma_z+\sigma_z v_y)
\\ =&&\frac{1}{2}\frac{p_y}{m^*}\sigma_z.
\end{eqnarray}
On the basis of Kubo's formalism, the spin Hall conductivity in
response to a dc electric field at zero temperature can be expressed
as
\begin{eqnarray}\label{SHE}
\sigma_{sH}=-\frac{e}{2\pi}\mathrm{Tr}\bigl[J^z_y G^r(\mu) j_x
G^a(\mu)\big],
\end{eqnarray}
where
$j_x=\displaystyle\frac{d}{dp_x}\bigl(\frac{p^2}{2m^*}+\alpha(\sigma^x
p_y-\sigma^y p_x)\bigr)$ is the electrical current; $G^r$ and $G^a$
denote the retarded and advanced Green's functions, respectively.
The trace is taken over momentum and spin indices.

The unperturbated Hamiltonian $H_0=p^2/2m^*+\alpha(\sigma^x
p_y-\sigma^y p_x)$ can be diagonalized to be $\varepsilon_\pm (p)
=p^2/2m^*\mp p \alpha $ by the unitary matrix,
\begin{eqnarray*}
U(\textbf{p})= \frac{1}{\sqrt{2}}
\left( \begin{array}{cc} 1 & 1 \\
ie^{i\varphi_{\textbf{p} }} & -ie^{i\varphi_{\textbf{p} }}
\end{array} \right),
\end{eqnarray*}
where $\varphi_{\textbf{p}}$ refers to the azimuthal angle of the momentum
$\textbf{p}$ with respect to the $x$ axis.
Then the free particle Green's function in chiral bases can be expressed
as
\begin{eqnarray}
G_{0(ch)}^r(\textbf{p},\mu)= \left(\begin{array}{cc}
      \displaystyle \frac{1}{\mu-\varepsilon_+ +i \eta}  & 0\\[2mm]
      0   &  \displaystyle \frac{1}{\mu-\varepsilon_- +i \eta}
     \end{array}
 \right) .
\end{eqnarray}

We employ the diagrammatic technique~\cite{Pei} to calculate the
average spin Hall conductivity over the distribution of
impurities. The trace in Eq.(\ref{SHE}) is expanded as a sum of
diagrams:
\begin{widetext}\vspace{1mm}
\begin{center}
\begin{picture}(0,66)(0,0)
\put(-210,30){$\textrm{Tr}[J_y^z G^r j_x G^a]=$}
\put(-110,30){\begin{picture}(0,0) \put(0,0){\circle*{5}}
\put(60,0){\circle{5}} \qbezier(0,3)(0,25)(30,25)
\qbezier(60,3)(60,25)(30,25) \qbezier(0,-3)(0,-25)(30,-25)
\qbezier(60,-3)(60,-25)(30,-25) \put(30,-37){$\bar{G}^r$}
\put(30,30){$\bar{G}^a$} \put(-15,0){$J^z_y$} \put(65,0){$j_x$}
\end{picture} }
\put(-35,30){$+$} \put(-5,30){\begin{picture}(0,0)
\put(0,0){\circle*{5}} \put(60,0){\circle{5}}
\qbezier(0,3)(0,25)(30,25) \qbezier(60,3)(60,25)(30,25)
\qbezier(0,-3)(0,-25)(30,-25) \qbezier(60,-3)(60,-25)(30,-25)
\put(30,-37){$\bar{G}^r$} \put(30,30){$\bar{G}^a$}
\put(-15,0){$J^z_y$} \put(65,0){$j_x$}
\multiput(30,0)(0,2.5){11}{\circle*{1}}
\multiput(30,0)(0,-2.5){11}{\circle*{1}}
\end{picture} }
\put(75,30){$+$} \put(105,30){\begin{picture}(0,0)
\put(0,0){\circle*{5}} \put(60,0){\circle{5}}
\qbezier(0,3)(0,25)(30,25) \qbezier(60,3)(60,25)(30,25)
\qbezier(0,-3)(0,-25)(30,-25) \qbezier(60,-3)(60,-25)(30,-25)
\put(30,-37){$\bar{G}^r$} \put(30,30){$\bar{G}^a$}
\put(-15,0){$J^z_y$} \put(65,0){$j_x$}
\multiput(18,0)(0,2.5){10}{\circle*{1}}
\multiput(18,0)(0,-2.5){10}{\circle*{1}}
\multiput(42,0)(0,2.5){10}{\circle*{1}}
\multiput(42,0)(0,-2.5){10}{\circle*{1}}
\end{picture} }
\put(185,30){$+\cdots$ .}
\end{picture}\vspace{1mm}
\end{center}
\end{widetext}
The spin Hall conductivity consists of two parts $\sigma_{sH}^0$ and
$\sigma_{sH}^L$, namely,
\begin{eqnarray}\nonumber
\sigma_{sH}^0=-\frac{e}{2\pi}Tr\bigl[J^z_y \bar{G}^r(\mu) j_x \bar{G}^a(\mu)\big],\\
\sigma_{sH}^L=-\frac{e}{2\pi}Tr\bigl[\tilde{J}^z_y \bar{G}^r(\mu) j_x \bar{G}^a(\mu)
   \bigr],
\end{eqnarray}
where the former represents the contribution of one-loop diagram
while the latter arises from the vertex corrections. Here
$\tilde{J}^z_y$ refers to the corrected-spin-current vertex which
obeys a self-consistent equation~\cite{Pei} illustrated by the
following diagram,
\begin{center}
\begin{picture}(0,70)(0,0)
\put(-80,20){$\widetilde{J}_y^z=$} \put(-50,23){\begin{picture}(0,0)
\put(0,0){\circle*{5}} \put(0,0){\line(5,2){20}}
\put(0,0){\line(5,-2){20}} \multiput(20,-8)(0,2.5){7}{\circle*{1}}
\end{picture} }
\put(-20,20){+} \put(0,23){\begin{picture}(0,0)
\put(0,0){\circle*{5}} \put(0,0){\line(5,2){40}}
\put(0,0){\line(5,-2){40}} \multiput(20,-8)(0,2.5){7}{\circle*{1}}
\multiput(40,-15)(0,2.5){13}{\circle*{1}}
\end{picture}}
\put(50,20){$+\cdots$}
\end{picture}
\end{center}
Solving the self-consistent Born's equation, one obtain the
Green's function in the presence of impurities:
\begin{eqnarray}\label{eq:nodephase}
\bar G_{(ch)}^r(\textbf{p},\mu)= \left(\begin{array}{cc}
      \displaystyle \frac{1}{\mu-\varepsilon_+ + \displaystyle \frac{i}{2\tau }}  & 0\\[2mm]
      0   &  \displaystyle \frac{1}{\mu-\varepsilon_- + \displaystyle \frac{i}{2\tau }}
     \end{array}
 \right),
\end{eqnarray}
where the momentum-relaxation time $\tau$ is related to the
impurity concentration $n^{}_\mathrm{im}$
and scattering strength $u$,
namely $1/\tau=n^{}_\mathrm{im} u^2 m^*$ whatever for
nonmagnetic or magnetic impurities.

To uncover the puzzle and disentwine the controversy in the
impurity effects on the spin Hall conductivity, we need look
through the features of an infinite 2-dimensional quantum system.
An infinite system with infinitesimal impurity concentration
contains infinite number of impurities. Diluting the impurity
concentration means to increase the distance between impurities.
The infinite quantum systems with different impurity
concentrations can be mapped into each other by redefining the
length scale and Fermi wavelength, whereas they cannot be directly
mapped into a clean system. This means that the clean limit of an
infinite quantum system is actually not a clean system if the
effect of dephasing is ignored! The dephasing can be characterized
by inelastic relaxation time.

We employ an imaginary self-energy to represent the inelastic
scattering, which was early introduced in the weak localization
theory~\cite{Altshuler,Bergmann}. Thus the Green's functions $\bar
G^r(\mu)$ and $\bar G^a(\mu)$ are substituted by
frequency-dependent functions $\bar G^r(\mu+\omega/2)$ and $\bar
G^a(\mu-\omega/2)$ where $\omega$ is replaced by $i/\tau_i$ with
$\tau_i$ being the inelastic relaxation time (inelastic lifetime
in brevity). As a result, the averaged Green's function in chiral
bases turns to be
\begin{eqnarray}
\bar G_{(ch)}^r= \left(\begin{array}{cc}
      \displaystyle \frac{1}{\mu-\varepsilon_+ + \displaystyle \frac{i}{2\tau } + \displaystyle \frac{i}{2\tau_i } }  & 0\\[2mm]
      0   &  \displaystyle \frac{1}{\mu-\varepsilon_- + \displaystyle \frac{i}{2\tau } + \displaystyle \frac{i}{2\tau_i }}
     \end{array}
 \right).
\nonumber
 \\
\end{eqnarray}
In the limit case $\tau_i\rightarrow\infty$, the above Green's
function gives rise to Eq.~(\ref{eq:nodephase}).

\section{For nonmagnetic impurities}\label{sec:nonmagetic}

We consider a system in the presence of nonmagnetic impurities.
The interaction between electron and impurities is expressed as
\begin{eqnarray}\label{model}
V_{dis}=&& \sum_{i=1}^N \int d\textbf{r}^2 u
\delta(\textbf{r}-\textbf{R}_i) \hat{\psi}^\dag(\textbf{r})
\hat{\psi}(\textbf{r}).
\end{eqnarray}
The calculation of the momentum-integral of the product of Green's
functions is carried out analytically. For simplicity, we adopt
the limit of large Fermi energy $\mu \rightarrow \infty$ firstly.
The $\sigma_{sH}^0$ we obtained remains the so called universal
value $e/8\pi$.

Furthermore, we calculate the vertex correction.
The self-consistent equation for $\tilde J_y^z$ reads
\begin{eqnarray}
\widetilde{J}^z_y= \frac{n_i u^2}{V} \sum_{\textbf{p}} \bar{G}^a
(\textbf{p},0) (J^z_y(\textbf{p})+ \widetilde{J}^z_y )
\bar{G}^r(\textbf{p},0). \label{vertex1}
\end{eqnarray}
Solving the corrected-spin-current vertex, we get the corresponding matrix element, saying
\begin{eqnarray}
(\tilde{J}^z_y)_{\uparrow\downarrow}\displaystyle=\frac{-i}{4\alpha m^*
(2\tau-\tau')},
\end{eqnarray}
where $1/\tau'=1/\tau+1/\tau_i$.
Then we obtain the vertex correction
\begin{eqnarray}
\sigma_{sH}^L =
   \frac{-ie(\tilde{J}^z_y)^{}_{\uparrow\downarrow}
m^*\alpha\tau'}{2\pi }=\frac{-e}{8\pi
(2\displaystyle\frac{\tau}{\tau'}-1)}.
\end{eqnarray}
As a reasonable asymptotic behavior, the corrected-spin-current
vertex tends to zero when $\tau\rightarrow \infty$ which is the
realistic clean limit. In our formulation, the momentum-integral
of the product of Green's functions
$\sum_\mathbf{p}\bar{G}^r(\mu,\mathbf{p})j_x(\mathbf{p})\bar{G}^a(\mu,\mathbf{p})$
is convergent when $\tau\rightarrow \infty$ for a finite $\tau_i$
such that the vertex correction goes to zero in the clean limit.
It is due to the divergence of the momentum-integral when
$1/\tau_i =0$ in current literatures that the vertex correction is
not zero in the clean limit. The vertex correction arises from the
interference of the scattering waves from different impurities.
When the elastic lifetime is much larger than the inelastic
lifetime or the distance between impurities is much larger than
the decoherence length, the interference effect will disappear. So
it is natural that the vertex correction goes to zero in the clean
limit, removing the discontinuity in the conductivity variation.

The total spin Hall conductivity is then given by
\begin{eqnarray}
\sigma_{sH}=\frac{e}{8\pi}\frac{1}{
1+\displaystyle\frac{\tau_i}{2\tau}},
\end{eqnarray}
which fulfils
$\displaystyle\lim_{\tau\rightarrow\infty}\sigma_{sH}=e/8\pi$
for a finite inelastic lifetime.
The clean limit of the spin Hall conductivity is precisely the conductivity
in a clean 2DEG due to the disappearance of vertex corrections.
Here we do not consider the finite-size effect
since the system's size is assumed to be much larger
than the mean-free-path or the decoherence length.

In the limit $\tau_i\rightarrow\infty$,
we have $\sigma_{sH}=0$
in coincidence with the result for a system without inelastic scattering.
For large $\tau/\tau_i$, the spin transport belongs to the
semiclassical regime and the conductivity goes to $e/8\pi$.
When $\tau/\tau_i$ is small, the transport falls into the quantum regime and
the conductivity tends to zero.
Our result demonstrates the difference between the semiclassical
and quantum transport regimes in the spin Hall effect.
It was proved that the spin Hall conductivity vanishes
when there exist nonmagnetic impurities
in a homogenous system~\cite{Rashba04,Dimitrova}.
It was also indicated~\cite{Rashba04} that inhomogeneities facilitate
spin currents.
However, their conclusions are not valid after taking account of
inelastic lifetime because their Schr\"odinger equation is not
appropriate to describe a system with inelastic scattering.
Our consequence is that the spin Hall conductivity is not zero in a realistic system.

In calculating the momentum-integral of Green's functions in the
above, we assumed an infinite Fermi energy for simplicity. For a
finite Fermi energy, we can also evaluate the spin Hall
conductivity in semiclassical approximation~\cite{Dimitrova,Pei}.
We get formally a similar $\sigma_{sH}^0$ as in
Ref.~\cite{Dimitrova} with the only difference that $\tau$ becomes
$\tau'$,
\begin{eqnarray}\label{sigma0}
\sigma^0_{sH}=\frac{e}{8\pi}(1-\frac{1}{1+(\Delta \tau')^2}).
\end{eqnarray}
Now the matrix element of the corrected-spin-current vertex is found to be
\begin{eqnarray}
(\tilde J^z_y)^{}_{\uparrow \downarrow} =\frac{-iv_F \Delta \tau'^2}{4
\tau (1+(\Delta \tau')^2) - 2\tau'(1+(\Delta \tau')^2) -2 \tau'},
\end{eqnarray}
where $\Delta$ is the spin-orbit splitting at the Fermi surface,
and $v_F$ the Fermi velocity.
The vertex correction is obtained as
\begin{eqnarray}\nonumber
\sigma^L_{sH}= && \frac{e}{8\pi}\frac{(\Delta \tau')^2}{1+(\Delta
\tau')^2} \times  \\ && \nonumber \frac{-\Delta^2 \tau'^3}{2\tau
(1+(\Delta \tau')^2)
-\tau'(1+(\Delta \tau')^2)- \tau'}. \\
\end{eqnarray}
Consequently, the total spin Hall conductivity is given by
\begin{eqnarray}\label{shcn}
\sigma_{sH}=\frac{e}{8\pi}\frac{(\tau_i \Delta)^2}
{(1+\tau_i/\tau)^2+ (\tau_i \Delta)^2(\tau_i/2\tau +1)}.
\end{eqnarray}
The vertex correction still goes to zero in the clean limit,
leading to continuous change of the conductivity with respect
to impurity concentration.
For a finite Fermi energy, the spin Hall conductivity depends not only
on the ratio of elastic to inelastic lifetimes but also on the
spin-orbit splitting at the Fermi surface. The clean limit of Eq.
(\ref{shcn}) is $\displaystyle\frac{e}{8\pi} \frac{(\tau_i
\Delta)^2} {1+(\tau_i \Delta)^2}$, depending on the inelastic
lifetime $\tau_i$. If the inelastic lifetime becomes infinitely
long, Eq. (\ref{shcn}) diminishes to zero, recovering the result
we are familiar with.

Let us compare with the numerical result~\cite{Nomura051}.
Nomura et al. set a finite $\eta^{-1}$ to guarantee the convergence
in their numerical calculation and found the spin Hall conductivity
increases as the impurity concentration decreases and
$\eta\tau$ increases.
In their paper $\eta^{-1}$ is called the electric field
turn-on time, in fact, it should be the inelastic lifetime $\tau_i$.
Since $\eta \tau$ corresponds to the ratio $\tau/\tau_i$,
their result supports our present conclusion.

\section{For magnetic impurities}\label{sec:magnetic}

Very recently, the spin Hall conductivities in the presence of
magnetic impurities are calculated in Refs.~\cite{Inoue06,Pei}
manifesting the spin transport properties of magnetically and
non-magnetically disordered systems are different. It becomes an
obliged task to confirm whether this difference remains when the
inelastic scattering is not ignored.
We adopt the scattering potentials of magnetic impurities as
\begin{eqnarray}
V_{dis}=&& \sum_{i=1}^N \int d\textbf{r}^2 u \delta(\textbf{r}-\textbf{R}_i)\times
           \nonumber\\
&& \hat{\psi}^\dag(\textbf{r}) \left(
\begin{array}{cc}
  \cos\theta_i & \sin\theta_i e^{-i\phi_i} \\
  \sin\theta_i e^{i\phi_i} & -\cos\theta_i
\end{array}
\right) \hat{\psi}(\textbf{r}).
\end{eqnarray}
We firstly consider the infinite Fermi energy limit, obtaining
$\sigma_{sH}^{M0}= e/8\pi$. The self-consistent equation of
corrected vertex is expressed as
\begin{eqnarray}
\widetilde{J}^{Mz}_y= \frac{n_i u^2}{V} \sum_{\textbf{p}}
 \int d\theta d\phi  \frac{1}{4\pi} \sin\theta
\left( \begin{array}{cc} \cos\theta & \sin\theta e^{-i\phi} \\
      \sin\theta e^{i\phi} & -\cos\theta \end{array} \right)\times
         \nonumber\\
\bar{G}^a (\textbf{p},0) (J^z_y(\textbf{p})+
 \widetilde{J}^{Mz}_y ) \bar{G}^r(\textbf{p},0)
   \left( \begin{array}{cc} \cos\theta & \sin\theta e^{-i\phi} \\
\sin\theta e^{i\phi} & -\cos\theta \end{array} \right).
       \nonumber\\
\label{vertex1}
\end{eqnarray}
The matrix element of corrected vertex is found to be
\begin{eqnarray}
(\tilde{J}^{Mz}_y )^{}_{\uparrow\downarrow} \displaystyle=\frac{i}{4\alpha
m^* (6\tau+\tau')}.
\end{eqnarray}
And the vertex correction is obtained
\begin{eqnarray}
\sigma_{sH}^{ML}=\frac{e}{8\pi}
\frac{1}{6\displaystyle\frac{\tau}{\tau'}+1}.
\end{eqnarray}
Thus the spin Hall conductivity is given by
\begin{eqnarray}
\sigma_{sH}^M=\frac{e}{8\pi}
(1+\frac{1}{6\displaystyle\frac{\tau}{\tau_i}+7}).
\end{eqnarray}
Obviously, the clean limit, $\tau\rightarrow\infty$,
of the spin Hall conductivity in magnetically disordered system
is also the universal value $e/8\pi$~\cite{sinova04}.
In the limit of infinite inelastic lifetime,
$\sigma_{sH}^M$ goes to $e/7\pi$ in consistent with the recent
results~\cite{Inoue06,Pei} for magnetic impurities.

We also consider the case for a finite Fermi energy.
The $\sigma^{M0}_{sH}$ here keeps the same result as Eq.~(\ref{sigma0})
for nonmagnetic impurities.
While the matrix element of the corrected vertex reads
\begin{eqnarray}
(\tilde{J}^{Mz}_y)^{}_{\uparrow\downarrow}
 \!=\! \frac{i v_F \Delta
\tau'^2}{12 \tau (1+(\Delta \tau')^2) + 2\tau'(1+(\Delta \tau')^2)
+ 2\tau'}
\end{eqnarray}
and vertex correction is given by
\begin{eqnarray}\nonumber
\sigma_{sH}^{ML}=&& \frac{e}{8\pi}\frac{(\Delta
\tau')^2}{1+(\Delta \tau')^2} \times  \\ && \nonumber
\frac{\Delta^2 \tau'^3}{6\tau (1+(\Delta \tau')^2) + \tau'
(1+(\Delta \tau')^2) + \tau'}. \\
\end{eqnarray}
The total spin Hall conductivity in the presence of magnetic
impurities is then expressed as
\begin{eqnarray}\label{shcm1}
\sigma^M_{sH}=\frac{e}{8\pi}\frac{(\tau_i \Delta )^2}
{(1+\tau_i/\tau)^2+ (\tau_i \Delta )^2
\displaystyle\frac{7+6\tau/\tau_i}{8+6\tau/\tau_i}}.
\end{eqnarray}
When $\tau/\tau_i=0$, Eq. (\ref{shcm1}) becomes
$\displaystyle\frac{e}{8\pi}\frac{8(\Delta\tau)^2}{8+ 7 (\Delta\tau)^2}$
recovering the result of our recent paper~\cite{Pei}.
When $\tau/\tau_i\rightarrow \infty$, Eq. (\ref{shcm1}) reduces to
$\displaystyle\frac{e}{8\pi}\frac{(\tau_i \Delta )^2} {1+(\tau_i \Delta )^2}$
which is precisely the conductivity in the clean limit of
nonmagnetic impurities for finite Fermi energy.

\section{Discussion on the impurity effects}\label{sec:impurity}

\begin{figure}
\includegraphics[width=79mm]{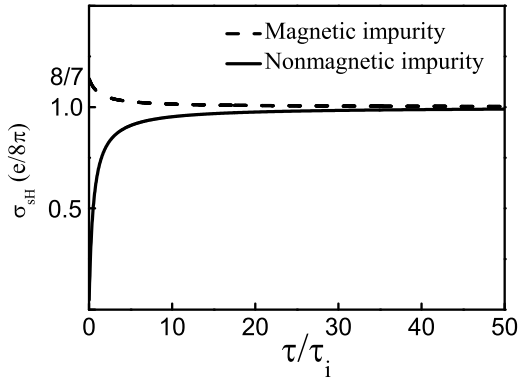}
\caption{\label{fig:conductivity}
Spin Hall conductivity versus the ratio of elastic to inelastic lifetimes
is plotted in both cases for nonmagnetic and magnetic impurities.}
\end{figure}

The above studies exhibited that
the spin Hall conductivities, in the presence of either nonmagnetic
or magnetic impurities, depend on the ratio of elastic to inelastic
lifetimes.
We plot the curves of the conductivity versus $\tau/\tau_i$ in the presence of
nonmagnetic and magnetic impurities respectively
in Fig.~\ref{fig:conductivity}.
As the impurity concentration decreases
(i.e., $\tau$ increases), $\sigma_{sH}^M$
decreases while $\sigma_{sH}$ increases monotonously.
The magnetic impurities enhance ISHE, while
the nonmagnetic impurities suppress it.
In the clean limit, both magnitudes approach to $e/8\pi$.
In the dirty limit or infinite inelastic-life-time limit
$\sigma_{sH}^M$ goes to $e/7\pi$ but $\sigma_{sH}$ goes to zero.

It is worthwhile to consider the finite-temperature case.
We have already shown that the spin Hall conductivity depends on the ratio of
elastic to inelastic lifetimes merely
in the approximation of infinitely wide band and infinitely large Fermi
energy.
These two characteristic times are important to both
the charge and spin transport properties for 2D systems.
Our results provide a clue to compare those characteristic times
by making use of the charge and spin transport experiments.
At low temperature, the elastic lifetime $\tau$ is determined by the
impurity concentration and is a constant independent of the
temperature,
while the inelastic lifetime $\tau_i$ relates to the electron-electron
interaction and decreases when the temperature increases.
In general, $\tau_i \propto T^{-p}$ whose exponent $p$ depends on
the scattering mechanism (e.g., $p=2$ for electron-electron scattering),
then the spin Hall conductivity is proportional to $1/(1+K T^{-p})$
increasing as the temperature raising.
Since the spin Hall conductivity is sensitive to the temperature,
its temperature dependence is expected to distinguish ISHE and ESHE.

\section{Spin Accumulation}\label{sec:accumulation}

The spin accumulation brought about by spin Hall effect has been
observed ~\cite{Kato,Sih,Nomura053,Kato05,Wunderlich,Valenzuela}
in experiments. We therefore evaluate the spin accumulation
generated by intrinsic spin Hall current based on the results we
obtained in the above. We consider a bar of width $W$ in
$y$-direction with an applied electric field along $x$-direction,
for which a spin current along $y$-direction will take place. If
the width is much larger than the mean-free-path and the
decoherence length $W\gg l, \, l_\varphi$, the system is in the
diffusive transport regime. The spin accumulation can be studied
by the diffusion equation
\begin{eqnarray}
D \frac{d^2 S_z}{dy^2} = \frac{S_z}{\tau_s},
\end{eqnarray}
where $D=v^2_F\tau/2$ is the diffusive coefficient,
$\tau_s$ the spin relaxation time (spin lifetime in brevity),
and $S_z =1/2 (n_\uparrow - n _\downarrow)$.
The spins accumulate at the edges of the bar until the
diffusive spin current in opposite direction balances the spin Hall current.
Then a steady spin distribution is established.
For simplicity, we do not take the influence of the spin accumulation
on the spin Hall current into account.
The boundary condition is then given by
\begin{eqnarray}
J_y^z = D \frac{d S_z}{dy} \Bigr|_{y= \pm W/2}.
\end{eqnarray}
The solution of the above diffusion equation is
\begin{eqnarray}\label{dis}
S_z(y)= \frac{J_y^z \sqrt{\tau_s/2D }}{\cosh \displaystyle \frac
{W}{\sqrt{2D\tau_s}} } \sinh \frac{y}{\sqrt {D\tau_s/2}}.
\end{eqnarray}
Eq. (\ref{dis}) is a good approximation
for the spin distribution in the region away from the edge.
Approaching the edge, however, the spin accumulation will affect
the spin current significantly, leading to a dramatic loss of spin
current.
According to Eq. (\ref{dis}), the spin density at the edge
of the sample is
\begin{eqnarray}
S_z= \frac{e E}{8\pi v_F} \frac{2\tau/\tau_i}{2\tau/\tau_i+1}
\sqrt{\frac{\tau_s}{\tau}} \tanh \frac{W}{v_F \sqrt {\tau\tau_s}}
\end{eqnarray}
in the presence of non-magnetic impurities, and
\begin{eqnarray}
S_z= \frac{e E}{8\pi v_F} \frac{6\tau/\tau_i+8}{6\tau/\tau_i+7}
\sqrt{\frac{\tau_s}{\tau}} \tanh \frac{W}{v_F \sqrt {\tau\tau_s}}
\end{eqnarray}
in the presence of magnetic impurities.
Here $E$ stands for the electric field in $x$-direction.
The spin accumulation depends on the three
characteristic times: the elastic lifetime, the
inelastic lifetime and the spin lifetime.
Because the different characteristic times can be manipulated separately,
our result proposes an alternate route to verify the ISHE
by measuring the spin Hall accumulation under different conditions.

The spin accumulation due to spin Hall effect was analyzed
theoretically~\cite{Rashba05,Nomura053,Sarma06,Sinova05}.
In 2DEG, the spin accumulation was observed in
Ref.~\cite{Kato05} where the authors regarded it arising from ESHE
because the relation $\Delta < 1/\tau$ is satisfied in their experiments.
In this situation, they believe the ISHE does not exist according to Nomura's
calculation~\cite{Nomura051}.
But our result shows that the spin Hall conductivity is not zero, instead,
it depends on the characteristic times of the system,
especially on the ratio of elastic to inelastic lifetimes.
Whether the spin accumulation observed in experiment comes from ESHE or
ISHE is worthy analyzed more carefully. An experiment at different temperature
is expected to be helpful.

\section{Conclusive Remarks}\label{sec:summary}

We indicated that the clean limit of an
infinite quantum system is not a clean system if the
effect of dephasing is ignored.
It was hence natural for one to have
obtained a discontinuity (zero with arbitrarily small impurity
concentration but $e/8\pi$ with no impurities) in the spin Hall
conductivity for infinite systems without taking account of
dephasing which is characterized by inelastic relaxation time.
The disagreements between numerical result and the others' analytical
results become inevitable because the former~\cite{Nomura051}
dealt with a finite system and the clean limit of a finite system
is a clean system.

We exposed the influence of inelastic relaxation
time to ISHE for 2D electrons in the presence of magnetic and
nonmagnetic impurities, respectively.
We found that the inelastic scattering plays an important role in the spin Hall effect,
leading to a nonzero spin Hall conductivity for arbitrary impurity concentrations.
In the dirty limit, the spin Hall conductivity goes to zero and $e/7\pi$
for nonmagnetic or magnetic impurities, respectively.
It tends to $e/8\pi$ in the clean limit
regardless of magnetically or non-magnetically disordered systems.
We revealed the importance of characteristic times,
such as the elastic, inelastic and spin lifetimes for ISHE.
The spin Hall conductivity is shown to depend
on the ratio of elastic to inelastic lifetime
and varies when temperature changes,
which provides a method to distinguish
the ISHE and ESHE by measuring the spin current at different temperatures.
Based on the spin Hall conductivity we obtained,
we evaluated the spin accumulation and presented an alternate route to
verify the ISHE by measuring it under different conditions.

\acknowledgements

The work was supported by Program for Changjiang Scholars and Innovative
Research Team in University, and NSFC Grant Nos. 10225419 and 10674117.

\end{document}